\def\kms{\relax \ifmmode {\,\rm km\,s}^{-1}\else \,km\,s$^{-1}$\fi}

\def\farcs{\hbox{$.\!\!^{\prime\prime}$}}

\def\arcmin{\hbox{$^\prime$}}
\def\arcsec{\hbox{$^{\prime\prime}$}}
\def\secd#1.#2{ #1\farcs#2 }               % seconds over decimal point

\def\mincir{\ \raise-2.truept\hbox{\rlap{\hbox{$\sim$}}\raise5.truept
    \hbox{$<$}\ }}
\def\magcir{\ \raise-2.truept\hbox{\rlap{\hbox{$\sim$}}\raise5.truept
    \hbox{$>$}\ }}

\def\nii{[N~{\sc ii}]}
\def\oiii{[O~{\sc iii}]}
\def\ha{H$\alpha$}
\def\stry{{\it Str\"om.\ Y}}
\def\hii{H~{\sc ii}}

\documentclass{aa}  
\usepackage[dvips]{graphics}
\begin{document}
%\thesaurus{8(09.16.1; 11.09.1 M33; 11.09.4)}
\title{New candidate planetary nebulae in M~81\thanks{Based on 
observations obtained at the 2.5m~INT telescope operated on the island
of La Palma by the Isaac Newton Group in the Spanish Observatorio del
Roque de Los Muchachos of the Instituto de Astrofisica de Canarias.}}
\author{Laura Magrini   
\inst{1}, 
Mario Perinotto
\inst{1},
Romano~L.M. Corradi
\inst{2},
Antonio Mampaso
\inst{3}}
\offprints{M.~Perinotto\\ 
e-mail: mariop@arcetri.astro.it}
\institute{Dipartimento di Astronomia e Scienza dello Spazio,
Universit\'a di Firenze, L.go E. Fermi 5, I-50125 Firenze, Italy 
\and 
Isaac Newton Group of Telescopes, Apartado de Correos 321, 38700 Santa 
Cruz de La Palma, Canarias, Spain
\and
Instituto de Astrof\'{\i}sica de Canarias, c. V\'{\i}a L\'actea s/n, 
38200, La Laguna, Tenerife, Canarias, Spain} 
\date{received date; accepted date}
\authorrunning{Magrini et al.}
\titlerunning{New candidate planetary nebulae in M~81}
\maketitle
\abstract A 34\arcmin$\times$34\arcmin\ field centred on the spiral galaxy M~81
has been searched for emission-line objects using the prime focus wide
field camera (WFC) of the 2.54~m Isaac Newton Telescope (La Palma,
Spain). A total of 171 candidate planetary nebulae (PNe) are found, 54
of which are in common with the ones detected by Jacoby et al. (1989). 
The behaviour of PNe excitation as a function of galactocentric
distance is examined, and no significant variations are found.
 The PNe luminosity function is built for the disk and bulge of M~81, separately.  A
distance modulus of $27.92 \pm 0.23$~mag is found for disk PNe, in
good agreement with previous distance measurements for M~81 (Jacoby et
al. 1989; Huterer et al. 1995).


\section{Introduction}
Basic information about planetary nebulae (PNe) in external galaxies
has been summarized by Jacoby (1997). Many hundreds of PNe were discovered in
spiral galaxies (mainly in M~31), in irregular (LMC and SMC) and
elliptical galaxies up to the distance of the Virgo cluster. More
recently, 131 candidate PNe have been discovered by Magrini et
al. (2000, 2001) in M~33.  PNe are important objects in the study of stellar
population of intermediate age in galaxies of different morphological
types and in different chemical environments, and in the assessment of the
kinematical properties of all morphological components of galaxies
(disks, bulges, haloes).

M~81 (NGC~3031), a nearby galaxy outside the Local Group, belongs to
the group of the most massive spiral galaxies (SAab) being also 
the nearest LINER galaxy. Together with M~82 (NGC~3034), NGC~3077 and
various dwarf galaxies, it forms a quite interesting multiple system.
The central area of 4\arcmin$\times$4\arcmin\ of M~81 was searched for
PNe by Jacoby et al. (1989), who discovered 185 candidate PNe.  In the
present paper we address the search of PNe in a much more extended
region, a factor of 70 times larger than the already surveyed area,
albeit with a 0.7 mag lower detection limit in the central
crowded regions.  Our survey covers the whole optical extent of the
galaxy and its surroundings.  In Sect.~2 observations and data
reduction are presented. Sect.~3 contains the data analysis. In
Sect.~4 we introduce the newly discovered candidate PNe, while in
Sect.~5 we discuss the behaviour of their excitation in
comparison with other galaxies.  In Sect.~6 we build the PNe
luminosity function in order to evaluate the completeness of our
sample and to use it as a standard candle for extragalactic distance
determination, thus deriving the corresponding distance modulus of M~81.
The summary and conclusions are presented in Sect.~7.

\section{Observations and data reduction}

The M~81 galaxy was observed on December 14, 2000 and January 2, 2001
using the prime focus wide field camera (WFC) of the 2.54~m Isaac
Newton Telescope (La Palma, Spain).  An area of
34\arcmin$\times$34\arcmin, covering the whole galaxy (see Fig.~1),
was surveyed with a panoramic detector consisting of four EEV CCD of 
 $2048 \times 4096$ pixels each.  The pixel size projects to
0$''$.33 in the sky. Observations were taken through three filters
with the following central wavelengths and widths 
(FWHM): an \oiii\, filter (5008/100~\AA), an \ha+\nii\, filter
(6568/95~\AA) and a Str\"omgren~$Y$ filter (5550/300~\AA), 
the latter used as an
off-band filter for identifying the emission-line sources.  The total
exposure through the \oiii\, filter was of 6600 s split into two
sub-exposures of 40 minutes and one of 30 minutes.  The \ha+\nii\,
image had a total exposure time of 1800 s (three 600 s frames) whereas
the Str\"omgren~$Y$ image was exposed 1600 s (four 400 s exposures).
The centre of the telescope pointing was: 09h 55m 33.2s, +69d 03m 55s
(J2000.0).  The seeing was approximately 1$''$.5 in the
\oiii\, and \stry\ frames, and 1$''$ in the \ha+\nii\, frames.  The
CCD frames were de-biased and flat-fielded using IRAF.  All images 
were aligned to a reference image, correcting for
geometrical distortion. Images in the same filter were then averaged
out to remove cosmic rays and to improve the signal to noise
ratio. Sky background was subtracted using the external regions of our
frames, where the optical emission of the galaxy is negligible.
 
\begin{figure*}
\resizebox{\hsize}{!}{\includegraphics{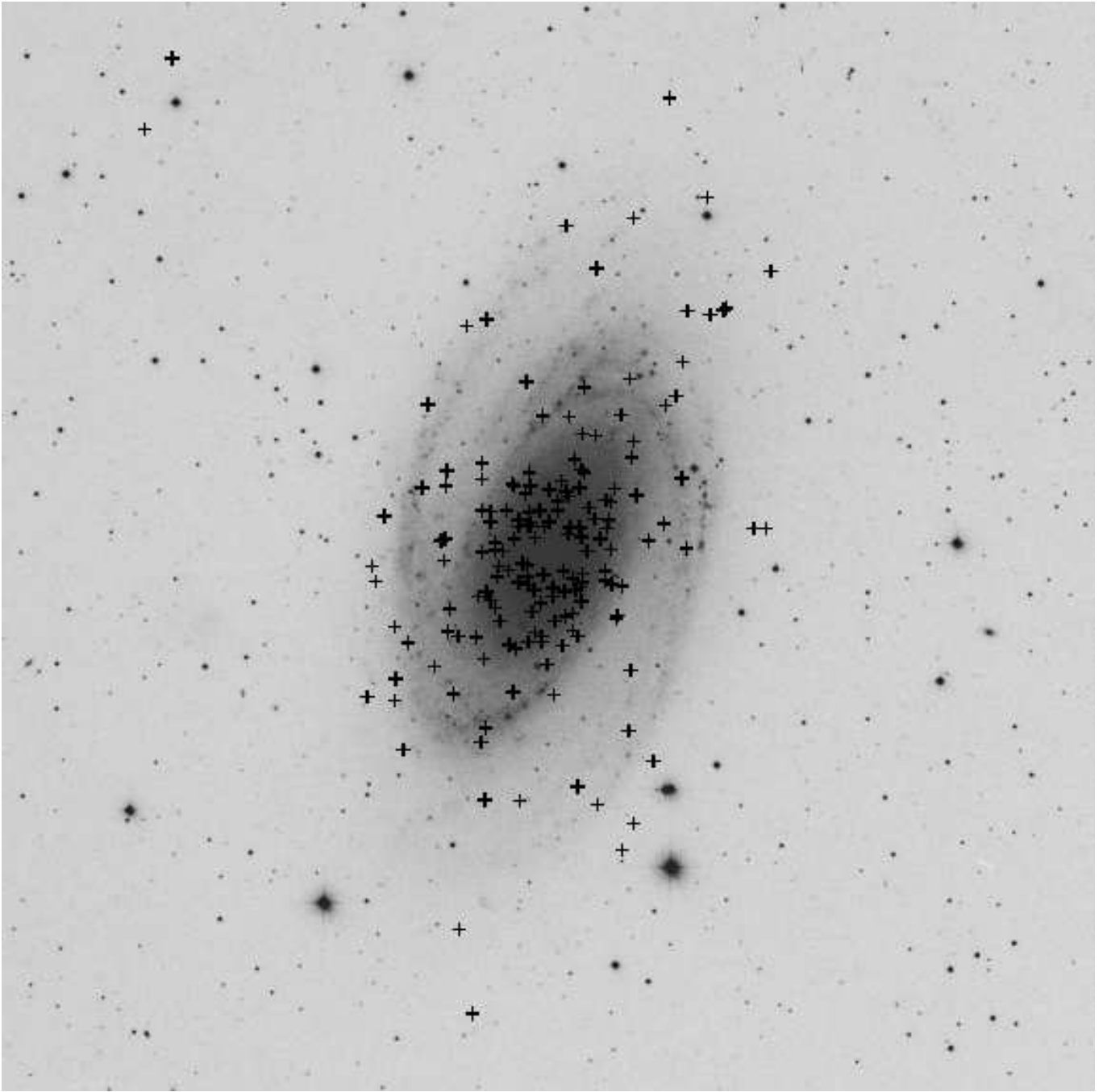}}
\caption{The Palomar Atlas image of M~81. The field of this frame is 
approximately 30\arcmin$\times$30\arcmin. The field of view of the
INT+WFC is slightly larger (34\arcmin$\times$34\arcmin), but we show
here only the area in which we have found PNe.  North is at the top, East
to the left. The ``$+$'' symbols indicate the location of the
candidate PNe.}
\label{m81}
\end{figure*} 

\section{Data analysis}

Emission-line objects were identified by subtracting the stellar
continuum of M~81 using the off-band, \stry\, frames from the \oiii\,
and \ha+\nii\, frames, as described in Magrini et al. (2000).  This
procedure removes the stellar background allowing for identification of
emission-line sources.

Following Jacoby et al. (1989), who tested the differences between
crowded field photometry versus aperture photometry for the M~81 bulge
PNe, we adopted the aperture photometry technique. \oiii\
and \ha+\nii\ line fluxes were then measured using APPHOT in the
continuum-subtracted images. Errors were estimated considering both
photometric errors, given by Poissonian statistics on the background
and on the sources, and the detector noise. Photometric errors vary
between few per cent for the brightest objects in the \oiii\, and \ha
+ \nii\, images up to about 20$\%$ for the fainter ones in the \oiii\,
frames, whereas they can reach 50-60$\%$ for the faintest candidate
PNe in the \ha+\nii\, images, where the signal to noise ratio is lower
due to the shorter exposures.  Instrumental fluxes were transformed
into physical fluxes using the standard PN calibrator 
PNG~205.8-26.7, whose accurate fluxes are given in Dopita \& Hua
(1997).

Emission-line fluxes were then corrected for the interstellar
extinction. Following Jacoby et al. (1989), we consider only the
foreground Galactic extinction towards M~81, which is quoted to be
$A_V\simeq 0.3$ mag, and corresponds to $E(B-V) \simeq 0.1$ mag (Kaufman et
al. 1987). We have adopted this value to correct the \oiii\ and
\ha+\nii\, fluxes according to Seaton (1979).

Coordinates of the emission-line sources in our images were derived
with a multi-step procedure. Firstly, approximately 30 stars for each
\oiii\ CCD field were identified on the Digitized Sky Survey \footnote{
Based on photographic data of the National Geographic Society Palomar
Observatory Sky Survey (NGS-POSS) obtained using the Oschin Telescope
on Palomar Mountain.}. A first astrometric solution was computed with
these stars, and with this solution coordinates for about 400 stars 
found with DAOFIND in the field were subsequently obtained. Their
coordinates were then replaced with the nearest APM-POSS1 coordinates and a
new astrometric solution was computed. The procedure was iterated
to reach a final accuracy of approximately 0.\arcsec5 r.m.s.

\section{Candidate planetary nebulae}
To identify PNe, we have adopted the following criteria:
\begin{itemize}
\item[i)] the object should appear both in the \oiii\, and \ha+\nii\, 
frames but not in the continuum frame. A negligible continuum is in fact
expected from a PN at these wavelengths;
\item[ii)] it should not be spatially resolved, i.e. it should have a 
stellar point spread function. 
At the distance to M~81 (3.50$\pm$0.40 Mpc, Jacoby et
al. 1989), 1\arcsec\ corresponds to approximately 17~pc whereas the
largest galactic PNe have sizes of $\sim$4~pc (cf. Peimbert 1990;
Corradi et al. 1997). Therefore, all PNe belonging to M~81 are
expected to be unresolved in our images, and any extended emission
region is considered to be an \hii\ region or a SN remnant.
\end{itemize} 
One hundred and seventy-one objects fulfilling criteria i) and ii) have been 
found. They are listed in Table 1 together with their observed \oiii\,
and \ha+\nii\, fluxes and coordinates; they are indicated in Fig.~1 by
``+'' signs.  Fifty four candidates coincide with the objects
discovered by Jacoby at al. (1989) in the bulge of M~81; their
coordinates and fluxes agree with those in Jacoby at al. (1989) within
the errors.  In column 6 of Tab.~1 the identification number in
Jacoby et al. (1989) is reported.

Magrini et al. (2000) have shown that the $R$=$I$(\oiii)/$I$(\ha+\nii)
flux ratio can be used to distinguish PNe from other emission-line
objects in a statistical sense, but not for individual objects.  In
fact, the analysis of the catalogue of Galactic PNe by Acker et
al. (1992) shows that $R$ spans a wide range, roughly from 0 (\oiii\
undetected) to 10, and that $\sim$75$\%$ of PNe have $R>1$. \hii\
regions are instead generally of lower excitation ($R<1$), but this is
not always true, so that misclassification between PNe and compact
\hii\ regions (which appear as point-like at the distance of M~81) can
occur. Only detailed spectroscopy would allow a final distinction between the two classes of objects. Nevertheless, using the $R$ flux ratio it
is at least possible to estimate the contamination of \hii\ regions in
our sample in a statistical way, as done by Magrini et al. (2000).
For the 135 objects in Table~1 in which both the \oiii\
and \ha+\nii\, fluxes have been measured, we find 28$\%$ with $R<1$,
as compared to 25$\%$ for the Galactic PNe (808 PNe whose fluxes are quoted in
Acker et al. 1992).  Assuming that M~81 and the Milky Way contain the same 
percentage of low excitation PNe, then we expect only
a few \hii\ regions (approximatively 5) contaminating our sample of
candidate PNe.

\section{Radial distribution of excitation and comparison 
with other galaxies}

The excitation class of a PN is defined in terms of flux ratios,
primarily of \oiii\, to \ha\, or H$\beta$ but also considering other
nebular lines (Feast 1968). These ratios are good 
indicators of the temperature of the central star but also depend on the
properties of the nebula like its geometry, electron density and
temperature (and therefore the abundance of the important coolant O/H)
and on the optical depth in the Lyman continuum.  The flux ratio
$R=\frac{[OIII]}{H\alpha+[NII]}$ thus provides us with an indication
of the excitation of the candidate PNe in M~81, as already anticipated
in the previous section.

\begin{figure}
\resizebox{\hsize}{!}{\includegraphics{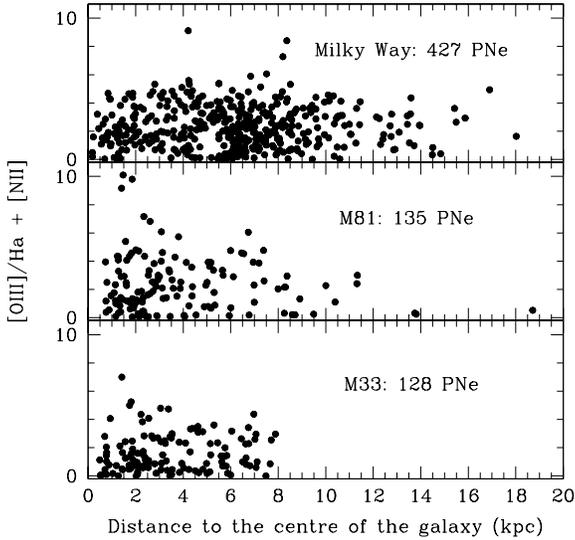}}
\caption{The flux ratio $R=\frac{[OIII]}{H\alpha+[NII]}$ as a 
function of the galactocentric distance for the Milky Way (data from
Acker et al. 1992, upper box), M~81 (this work, middle box), and M~33
(Magrini et al. 2000, Magrini et al. 2001, lower box).}
\label{m81m33mw}
\end{figure}

In the Galaxy, Vorontsov-Vel'yaminov et al. (1975) presented an absolute
spectrophotometry of 47 PNe in the direction of the Galactic centre.
They found that their $R$ ratios differ significantly from those of
disk PNe, concluding that there are physical differences between
Galactic bulge and disk PNe.  Webster (1975, 1976) compared the
distribution of excitation of the Galactic bulge PNe with those in the
Magellanic Clouds and proposed the presence of more massive
progenitors in the case of LMC, where there are more high excitation
PNe than low excitation ones, compared to our Galaxy and the SMC. 
In spiral galaxies, information on the excitation of PNe
thorough the bulge and the disk is available only for M~81 (this
paper), M~33 (Magrini et al. 2000) and the Galaxy (Acker et
al. 1992). Using the information contained in the Strasbourg ESO
Catalogues, the galactocentric distances of 427 Galactic PNe 
with quoted \ha, \nii\, and \oiii\, fluxes  were
computed; the distances from the Sun of the PNe come from
Cahn et al. (1992), and the adopted galactocentric distance of the Sun 
is 8.5$\pm$1.1 kpc (Allen, 2000).  

Fig.~\ref{m81m33mw} shows $R$ as a function of the distance from the
centre of the galaxy in the three cases. PNe in M~33 and the Milky Way
are approximately homogeneously distributed in terms of excitation.  
The apparent progressive reduction of high excitation PNe with galactocentric
distance in M~81 (Fig. 2) is removed
if the mean $R$, computed in bins of 2~kpc, is plotted (Fig.~3).

Therefore, and contrary to Vorontsov-Vel'yaminov et al. (1975), no significant 
difference between the excitation of bulge and disk PNe is found in the 
Galaxy nor in M~81 (Fig.~\ref{vvhisto}). 

\begin{figure}
\resizebox{\hsize}{!}{\includegraphics{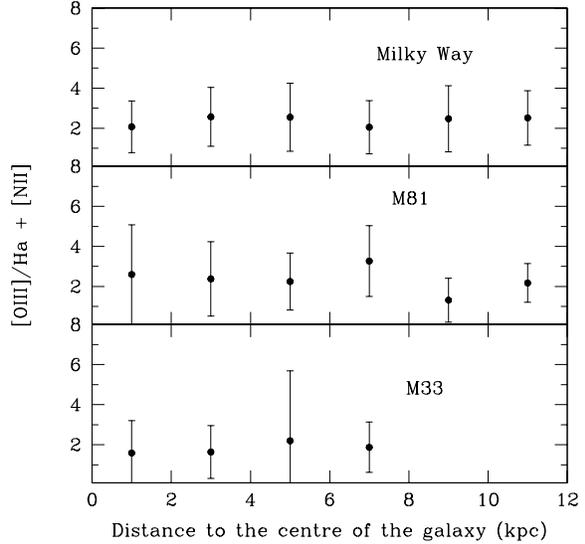}}
\caption{As in Fig.2, but plotting the mean value of $R$, averaged on bins
of 2 kpc and up to the distance of 12 kpc. 
The error bars indicate the standard deviation from the mean.}
\label{radhisto}
\end{figure}

\begin{figure}
\resizebox{\hsize}{!}{\includegraphics{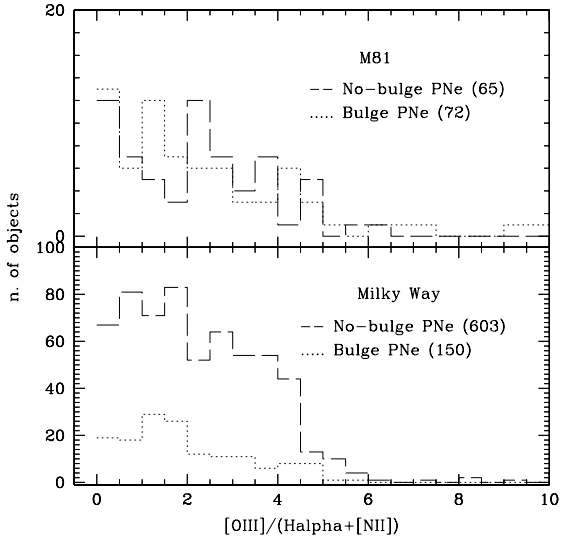}}
\caption{Histograms of R for M~81 (this work, upper box) and the Milky Way (lower box),
for disk and bulge PNe separately.}
\label{vvhisto}
\end{figure}

\section{Planetary Nebulae Luminosity Function}

\begin{figure}
\resizebox{\hsize}{!}{\includegraphics{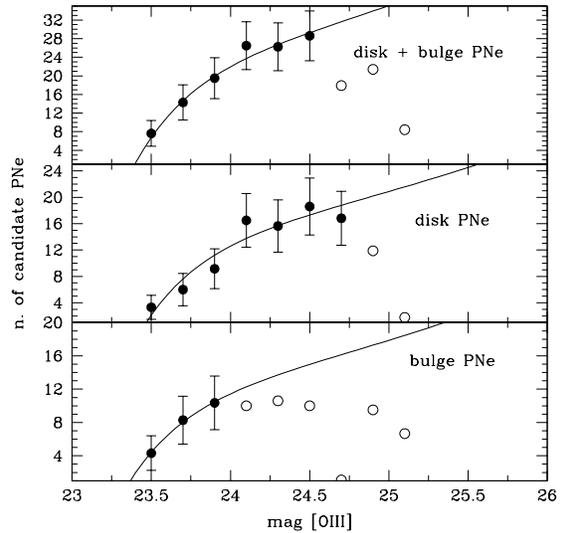}}
\caption{Mean square fits for the luminosity functions of PNe in M~81: i) Complete sample 
(171 candidate PNe); ii) Disk (86 candidate PNe); iii) Bulge and
central regions (85 candidate PNe). Black points indicate candidate PNe within the completeness limits.  }
\label{pnlf}
\end{figure} 

\oiii\ fluxes for the candidate PNe in M~81 were converted into 
equivalent V-band magnitudes following Jacoby (1989):
\begin{equation}
m_{\rm [O~{III}]}=-2.5\log F_{\rm [O~{III}]} -13.74.
\nonumber
\end{equation}

The Planetary Nebulae luminosity function (PNLF) was then built for
three different samples: all 171 PNe of M~81, those in the disk (86 objects)
and those in the bulge (85). They are presented in
Fig.~\ref{pnlf}. The Eddington formula (Eddington 1913) was applied to
correct for the effects of observational errors.  The resulting
luminosity function within the completeness limit was then fit to the
``universal'' PNLF
\begin{equation}
N(m)\propto e^{0.307(m)}(1-e^{3(m_{\ast}-m)}).
\label{pnlffit}
\nonumber
\end{equation}
In Eq. (2) $m$ is the observed \oiii\ magnitude from Eq. (1), and
$m_{\ast}$ is the apparent magnitude of the PNLF cutoff of M31 (Jacoby
1989).  As we have discussed in Magrini et al. (2000), it is possible
to estimate the limit of completeness of a sample of PNe by computing
the S/N ratio for each PNe (putting the limit of completeness at
S/N$\sim$10, Ciardullo et al. (1987)) or equivalently by evaluating
the point at which observational PNLFs inflect downwards indicating
incompleteness.  Fig.~\ref{pnlf} shows that  
the completeness limit changes for the different samples,
indicating the strong dependence on the galaxian background
emission, which increases towards the centre of the galaxy.  Comparing
our data with the sample of Jacoby et al. (1989), we note in fact that
we are losing approximately 70$\%$ of the PNe found by Jacoby and
collaborators in the bulge of M~81. This is due to two effects: the
use of a filter that is three times broader than the one used by
Jacoby et al. (1989), and a smaller primary mirror (2.5 m vs. 4 m) not
compensated by longer exposure times (1.8 hr vs. 1.5 hr).  These
effects are particulary important in the central regions because of
the high surface brightness of the parent galaxy. In these regions our
completeness limit of 24.0 mag differs from that of the
sample of Jacoby et al. (24.7 mag) and from the one we have for our
data in the disk (24.7 mag).  For this reason the distance
modulus of M~81 is computed using only disk PNe, where we consider to
have a complete sample.  The error on the distance modulus is obtained
by combining the error associated with our best least mean square fits
($\pm 0.04$ mag) in quadrature with those associated with the
photometric zero point (0.03 mag) and with the adopted extinction (0.2
mag from Kaufman et al. 1987). We have also to consider two systematic
errors which come from the uncertain definition of the empirical PNLF
(0.05 mag) and the distance of the calibration galaxy M31 (0.10
mag). These errors lead to a total error of $\pm 0.23$ mag.  The
resulting distance modulus to M~81 is $27.92 \pm 0.23$~mag,
corresponding to $3.84 \pm 0.41$~Mpc.  This distance is in excellent agreement
with the value derived from the Cepheids method of $27.97
\pm 0.14$~mag (Huterer at al. 1995) and also in fair agreement with the value derived by
Jacoby et al. (1989) using the luminosity function for their bulge PNe candidates
($27.72 \pm 0.25$~mag).

\section{Summary and conclusions}

We imaged a 34\arcmin$\times$34\arcmin\ area around the spiral galaxy
M~81 through \oiii\, and \ha+\nii\, narrowband filters. In total, 
171 candidate
PNe are found, 117 of which are new while 54 coincide with those
discovered by Jacoby et al. (1989) in the central
4\arcmin$\times$4\arcmin\ bulge.

The behaviour of PNe excitation across the galaxy was examined,
finding no evidence for substantial differences in excitation between
bulge and disk PNe, nor for variations along the galaxian disk.  The
same result applies to the Milky Way and M~33, and is contrary to previous
suggestions from Vorontsov-Vel'yaminov et al. (1975).

The PNLF of the candidate PNe in the disk of M~81 provides a distance
determination of $3.84 \pm 0.41$~Mpc.

\begin{acknowledgements}

We are grateful to Peter Sorensen for taking the observations of M~81
in service mode.

\end{acknowledgements}

%%%%% TAB 1

\begin{table*}
\caption{PN candidates found in M81. Fluxes in \ha+\nii\ and in \oiii\ are in
units of 10$^{-16}$~erg~cm$^{-2}$~s$^{-1}$ and are corrected for
reddening (see text).}

%{\scriptsize
{\tiny
\begin{tabular}{cc}
\begin{tabular}{r r r r r c } 
\hline
id  &  \multicolumn{2}{c}{R.A. (2000.0) Dec.} & $F_{[OIII]}$ & $F_{H\alpha}$ 
& Jacoby id \\ 
\hline
1&  9 54 20.35 &69  03 53.7  & 2.9  &   0.64 &   \\ 
2&  9 54 24.35 &69  03 56.0  & 3.0  &   --  &\\ 
3&  9 54 26.54 &69 10 30.5  & 5.5  &   25. &   \\ 
4&  9 54 27.66 &69 13 55.0  & 5.3  &   2.2  &   \\ 
5&  9 54 31.12 &69 10 25.8  & 4.1  &   1.9  &   \\ 
6&  9 54 36.70 &69 17 00.5   & 5.6  &   22. &   \\ 
7&  9 54 38.52 &69 10 36.8  & 4.9  &   2.4  &   \\ 
%8&  9 54 42.05 &69  9 6.0   & 17. &   47. &   \\ 
8&  9 54 45.50 &69  08 06.6   & 7.6  &   1.6  &   \\ 
9& 9 54 46.52 &69  05 39.0  & 11. &   56. &   \\ 
10& 9 54 49.97 &69  04 10.5  & 6.3  &   2.5  &   \\ 
11& 9 54 53.15 &69 13 32.2  & 5.0  &   2.2  &   \\ 
12& 9 54 54.24 &69  04 21.0  & 6.7  &   2.5  &   \\ 
13& 9 54 59.90 &69  03 53.6  & 3.5  &   1.2  &   \\ 
14& 9 54 09.28  &69 11 31.0  & 4.0  &   3.6  &   \\ 
15& 9 55 01.01  &69 12 04.3   & 5.3  &   1.8  &   \\ 
16& 9 55 01.47  &69  06 54.1  & 4.3  &   1.1  &   \\ 
17& 9 55 10.09 &69  02 37.1  & 6.4  &   70. &   \\ 
18& 9 55 10.25 &69  00 05.3   & 3.7  &   2.4  &   \\ 
19& 9 55 11.18 &69  05 09.9   & 5.3  &   0.7 & 4   \\ 
20& 9 55 12.24 &69  04 37.8  & 5.0  &   3.9  & 5   \\ 
21& 9 55 12.47 &69  05 13.6  & 4.9  &   4.4  & 11   \\ 
22& 9 55 12.65 &69  01 46.2  & 7.5  &   3.4  &   \\ 
23& 9 55 12.73 &69  03 46.0  & 3.9  &   2.9  & 12  \\ 
24& 9 55 13.12 &69  04 22.7  & 5.2  &   --  &14 \\ 
25& 9 55 13.35 &68 58 16.8  & 7.1  &   48. &   \\ 
26& 9 55 13.39 &69  01 40.8  & 5.0  &   --  &\\ 
27& 9 55 13.58 &69  02 43.0  & 7.2  &   84. &   \\ 
28& 9 55 13.62 &69  07 10.3  & 8.7  &   --  &\\ 
29& 9 55 14.78 &68 55 30.0  & 12. &   38. &   \\ 
30& 9 55 15.65 &69  02 48.9  & 3.9  &   --  &19\\ 
31& 9 55 15.65 &69  04 06.3   & 6.7  &   2.7  &   \\ 
32& 9 55 15.70 &69  08 39.8  & 6.5  &   2.6  &   \\ 
33& 9 55 15.96 &69 13 31.7  & 6.1  &   24. &   \\ 
34& 9 55 16.93 &69  04 42.9  & 5.7  &   --  &23\\ 
35& 9 55 18.39 &69  05 04.2   & 5.6  &   9.0  &27   \\ 
36& 9 55 18.97 &69  06 08.0   & 6.0  &   1.8  &   \\ 
37& 9 55 19.41 &69  06 12.2  & 4.5  &   --  &\\ 
38& 9 55 19.47 &68 54 45.7  & 5.6  &   4.2  &   \\ 
39& 9 55 19.55 &69  03 54.6  & 5.2  &   --  &\\ 
40& 9 55 02.48  &69  05 17.0  & 7.3  &   1.2  &   \\ 
41& 9 55 02.58  &69  06 26.3  & 9.9  &   6.5  &   \\ 
42& 9 55 20.53 &69  03 45.9  & 9.7  &   99. &   \\ 
43& 9 55 20.80 &69  05 41.7  & 6.6  &   --  &30\\ 
44& 9 55 21.53 &69  06 33.0  & 8.5  &   3.7  &   \\ 
45& 9 55 21.81 &69  07 50.6  & 9.1  &   120. &   \\ 
46& 9 55 22.14 &69  04 13.0  & 9.3  &   --  &35\\ 
47& 9 55 22.40 &69  04 30.5  & 4.9  &   2.1  &   \\ 
48& 9 55 22.81 &69  03 05.8   & 9.4  &   2.3  &37   \\ 
49& 9 55 23.38 &69  02 50.9  & 11. &   1.2  & 39  \\ 
50& 9 55 24.04 &69  02 43.4  & 10. &   0.99 & 42  \\ 
51& 9 55 24.40 &69  02 17.9  & 4.6  &   --  &46\\ 
52& 9 55 24.48 &69  03 37.9  & 6.3  &   --  &45\\ 
53& 9 55 24.99 &69  05 01.2   & 8.3  &   9.8  &48   \\ 
54& 9 55 25.01 &69  05 28.4  & 9.4  &   --  &49\\ 
55& 9 55 25.43 &69  05 37.8  & 9.8  &   1.0  &50   \\ 
56& 9 55 25.52 &69  04 31.6  & 7.4  &   --  &52\\ 
57& 9 55 25.65 &69  02 39.6  & 4.4  &   3.4  &54   \\ 
%59& 9 55 25.87 &68 56 10.7  & 17. &   53. &   \\ 
58& 9 55 26.30 &69  04 20.3  & 9.7  &   25. &56   \\ 
59& 9 55 26.38 &69  02 59.8  & 4.2  &   --  &\\ 
60& 9 55 26.66 &69  05 56.1  & 8.6  &   --  &57\\ 
61& 9 55 26.84 &69  01 16.6  & 11. &   5.9  &   \\ 
62& 9 55 27.64 &69  01 57.7  & 5.1  &   --  &\\ 
63& 9 55 28.25 &69  01 26.5  & 6.3  &   93. &   \\ 
64& 9 55 28.56 &69  05 22.0  & 4.1  &   1.4  &67   \\ 
65& 9 55 29.13 &69  05 04.0   & 7.9  &   --  &71\\ 
66& 9 55 29.20 &69  03 15.2  & 8.1  &   --  &76\\ 
67& 9 55 29.60 &69  02 38.6  & 6.6  &   --  &79\\ 
68& 9 55 30.24 &69  01 53.7  & 6.7  &   5.9  &   \\ 
69& 9 55 30.45 &69  03 34.1  & 10. &   --  &\\ 
70& 9 55 30.62 &69  07 56.3  & 11. &   67. &   \\ 
71& 9 55 31.36 &69  05 44.4  & 5.5  &   1.2  &82   \\ 
72& 9 55 31.95 &68 56 47.5  & 6.2  &   33. &   \\ 
73& 9 55 32.14 &69  01 02.7   & 9.5  &   67. &   \\ 
74& 9 55 32.27 &69  04 46.6  & 6.3  &   --  &86\\ 
75& 9 55 33.34 &69  02 46.0  & 4.3  &   4.3  &94   \\ 
76& 9 55 33.61 &69  02 32.7  & 9.9  &   3.4  &   \\ 
77& 9 55 33.68 &69  01 48.1  & 5.2  &   1.1  &   \\ 
78& 9 55 34.02 &69  04 39.2  & 7.5  &   1.9  &97   \\ 
79& 9 55 34.58 &69  09 00.9   & 5.1  &   1.4  &   \\ 
80& 9 55 34.97 &69  05 08.6   & 6.1  &   1.4  &100   \\ 
81& 9 55 36.02 &69  03 13.0  & 7.3  &   6.2  &107   \\ 
82& 9 55 36.26 &69  02 45.1  & 13. &   --  &\\ 
83& 9 55 36.81 &68 59 35.3  & 10. &   2.8  &   \\ 
84& 9 55 37.13 &69  05 53.8  & 4.8  &   1.0  &111   \\ 
85& 9 55 37.17 &69  06 18.1  & 12. &   82. &   \\ 
 &&&&&\\
\end{tabular}

& \phantom{pip}

\begin{tabular}{r r r r r c c} 
\hline
id  &  \multicolumn{2}{c}{R.A. (2000.0) Dec.} & $F_{[OIII]}$ & $F_{H\alpha}$ 
& Jacoby id \\ 

\hline
86& 9 55 37.78 &69  00 31.7  & 7.8  &   2.4  &   \\ 
87& 9 55 37.98 &69  02 21.4  & 5.5  &   4.6  &   \\ 
88& 9 55 38.12 &69  04 47.3  & 3.9  &   2.5  &121   \\ 
89& 9 55 38.62 &69  04 42.1  & 7.1  &   --  &126\\ 
90& 9 55 38.63 &69  05 06.8   & 7.9  &   2.4  &124   \\ 
91& 9 55 38.65 &69  01 23.3  & 4.5  &   --  &\\ 
92& 9 55 39.04 &69  01 12.3  & 6.9  &   1.8  &   \\ 
93& 9 55 39.07 &69  05 42.1  & 13. &   250. &   \\ 
94& 9 55 39.51 &69  02 46.7  & 12. &   35. &   \\ 
95& 9 55 39.77  &69  04 43.1  & 7.2  &   --  &131\\ 
96& 9 55 40.31  &69  04 53.7  & 3.4  &   2.8  &133  \\ 
97& 9 55 40.70 &69  01 30.1  & 3.4  &   2.1  &  \\ 
98& 9 55 41.00 &69  03 32.1  & 10. &   4.0  &136  \\ 
99& 9 55 41.11 &69  02 08.9   & 4.2  &   6.0  &137  \\ 
100& 9 55 41.20 &69  02 56.3  & 9.2  &   7.9  &  \\ 
101& 9 55 41.69 &69  03 11.0  & 4.6  &   --  &139\\ 
102& 9 55 42.23 &69  04 54.3  & 8.2  &   1.9  &143   \\ 
103& 9 55 42.50 &69  04 42.6  & 6.0  &   --  &144\\ 
104& 9 55 42.58 &69  03 36.9  & 10. &   18. &   \\ 
105& 9 55 42.74 &69  05 56.8  & 5.5  &   --  &145\\ 
106& 9 55 43.21 &69  06 01.5   & 3.8  &   8.5  &   \\ 
107& 9 55 43.39 &69  01 16.1  & 6.3  &   --  &\\ 
108& 9 55 44.36 &69  03 03.4   & 5.5  &   --  &152\\ 
109& 9 55 44.50 &69  04 21.4  & 5.3  &   3.0  & 153  \\ 
110& 9 55 45.97 &69  05 13.6  & 5.7  &   5.1  & 158  \\ 
%113& 9 55 46.14 &69 10 59.4  & 17. &   62. &   \\ 
111& 9 55 46.25 &69  06 21.5  & 7.3  &   --  &\\ 
112& 9 55 47.62 &69  01 05.4   & 2.0  &   0.95 &   \\ 
113& 9 55 47.72 &69  03 27.7  & 7.0  &   17. & 164  \\ 
114& 9 55 48.42 &69  02 19.8  & 5.6  &   3.1  &169   \\ 
115& 9 55 48.61 &69  04 02.6   & 4.0  &   2.8  &   \\ 
116& 9 55 49.70 &69  01 13.5  & 9.2  &   2.3  &   \\ 
117& 9 55 50.11 &68 59 48.4  & 7.7  &   3.5  &   \\ 
118& 9 55 50.50 &69  03 34.2  & 5.4  &   1.0  &177   \\ 
119& 9 55 50.96 &69  04 18.8  & 6.1  &   1.5  &180   \\ 
120& 9 55 51.34 &69  04 04.9   & 6.6  &   2.9  &181   \\ 
121& 9 55 51.46 &69  05 14.9  & 12. &   44. &   \\ 
122& 9 55 51.56 &68 56 33.0  & 5.8  &   1.5  &   \\ 
123& 9 55 51.61 &69  03 18.4  & 7.0  &   1.6  &183   \\ 
124& 9 55 51.73 &69  04 56.3  & 2.9  &   3.4  &182   \\ 
125& 9 55 52.14 &69  01 57.0  & 6.1  &   3.6  &   \\ 
126& 9 55 52.49 &69  06 43.3  & 4.1  &   2.9  &   \\ 
127& 9 55 52.93 &69 10 52.1  & 4.6  &   4.2  &   \\ 
128& 9 55 53.02 &69  06 15.2  & 6.7  &   2.5  &   \\ 
129& 9 55 53.30 &69  02 21.6  & 10. &   2.3  &   \\ 
130& 9 55 54.29 &69  05 18.6  & 11. &   24. &   \\ 
131& 9 55 54.66 &69  02 37.5  & 5.3  &   3.7  &   \\ 
132& 9 55 55.62 &69  02 53.2  & 5.4  &   2.9  &   \\ 
133& 9 55 55.69 &69  04 03.6   & 4.3  &   --  &\\ 
134& 9 55 58.53 &69  02 45.8  & 8.4  &   2.8  &   \\ 
135& 9 55 58.78 &69  00 53.1  & 6.3  &   1.1  &   \\ 
136& 9 55 06.11  &68 57 18.5  & 4.6  &   0.76 &   \\ 
137& 9 55 07.29  &69 12 10.4  & 5.2  &   2.4  &   \\ 
138& 9 55 09.25  &69  05 32.1  & 7.1  &   2.8  &   \\ 
139& 9 56 00.46  &68 58 50.0  & 5.7  &   1.7  &   \\ 
140& 9 56 00.56  &69  01 33.9  & 6.1  &   1.8  &   \\ 
141& 9 56 01.77  &69  01 24.6  & 2.7  &   1.3  &   \\ 
142& 9 56 13.41 &69  06 10.6  & 5.3  &   6.3  &   \\ 
143& 9 56 14.30 &68 50 20.5  & 6.1  &   19. &   \\ 
144& 9 56 15.41 &69  00 48.8  & 3.2  &   4.1  &   \\ 
145& 9 56 15.92 &68 52 54.1  & 3.3  &   1.1  &   \\ 
146& 9 56 02.41  &68 58 24.8  & 5.8  &   2.0  &   \\ 
147& 9 56 20.74 &68 58 26.8  & 5.5  &   1.4  &   \\ 
148& 9 56 23.37 &69  01 35.2  & 5.8  &   7.3  &   \\ 
149& 9 56 27.09 &69  05 26.4  & 11. &   2.8  &   \\ 
150& 9 56 27.29 &69  02 51.4  & 4.6  &   2.0  &   \\ 
151& 9 56 27.30 &69  02 07.0   & 4.3  &   1.1  &   \\ 
152& 9 56 28.38 &68 58 25.6  & 9.9  &   3.8  &   \\ 
153& 9 56 28.70 &69  00 32.8  & 12. &   17. &   \\ 
154& 9 56 29.82 &68 59 55.3  & 5.5  &   1.2  &   \\ 
155& 9 56 03.09  &68 56 39.7  & 4.0  &   0.84 &   \\ 
156& 9 56 03.73  &69  08 27.0  & 3.4  &   1.9  &   \\ 
157& 9 56 31.99 &69  03 31.3  & 7.4  &   2.2  &   \\ 
158& 9 56 32.84 &69  03 58.8  & 6.5  &   7.3  &   \\ 
159& 9 56 38.50 &69  00 07.0   & 3.7  &   1.6  &   \\ 
160& 9 56 04.48  &69  06 36.0  & 6.0  &   2.9  &   \\ 
161& 9 56 05.45  &69  06 10.3  & 6.0  &   1.4  &   \\ 
162& 9 56 06.20  &69  03 54.3  & 4.8  &   1.6  &   \\ 
163& 9 56 06.53  &69  01 36.9  & 5.7  &   -- &   \\ 
164& 9 56 07.50  &69  04 33.6  & 6.0  &   1.3  &   \\ 
165& 9 56 08.31  &69  02 29.0  & 3.4  &   37. &   \\ 
166& 9 56 08.41  &69  08 37.8  & 5.1  &   1.7  &   \\ 
167& 9 56 08.54  &69  03 56.0  & 4.9  &   3.6  &   \\ 
168& 9 56 09.03  &69  04 32.2  & 4.9  &   --  &\\ 
169& 9 56 09.87  &68 59 56.0  & 5.6  &   8.3  &   \\ 
170& 9 56 09.89  &69  01 47.9  & 8.1  &   -- &   \\ 
171& 9 57 23.75 &69 19 41.6  & 9.5  &   18. &   \\

\end{tabular}

\\
\end{tabular}
}
\end{table*}

\end{document}